\documentclass[a4paper,11pt]{article}
\pdfoutput=1 
\usepackage{jinstpub} 
\usepackage{lineno}

\newcommand{\comment}[1]{}
\newcommand{\lifzns}{$\mbox{Li}^6\mbox{F:ZnS}$}
\newcommand{\bnzns}{$\mbox{BN}\mbox{:ZnS}$}
\newcommand{\unit}[2]{$#1\mbox{~#2}$}

\title{Optimising Scintillating Neutron Detectors for Cosmic Ray Soil Moisture Monitoring}

\author[a,c]{P. Stowell \note{Corresponding author.}}
\author[b]{S. Fargher}
\author[c]{C. Steer}
\author[b,c]{L.F. Thompson}

\affiliation[a]{University of Durham, UK}
\affiliation[b]{University of Sheffield, UK}
\affiliation[c]{Geoptic Infrastructure Investigations Limited, UK}
\emailAdd{john.p.stowell@durham.ac.uk}


\abstract{Cosmic Ray Neutron Sensing (CRNS) is a powerful technique that allows non-invasive monitoring of soil moisture on length scales well matched for agricultural applications. One factor limiting the use of the technique within industrial agriculture settings is the high initial cost of Helium-3 or BF3 tubes typically used for ground level neutron monitoring. This paper discusses the use of Geant4 to optimise an alternative scintillator based neutron detector that may be applicable for challenges where cost is a higher driving factor than temporal resolution.}

\begin{document}
\maketitle
\flushbottom

\section{Introduction}
\label{sec:introduction}

By 2030 it is expected that, due to climate change, up to 50\% of the world's population will be living in an at-risk water supply area \cite{greve2018global}.  Approximately 69\% of fresh water withdrawal worldwide is  for irrigation purposes \cite{connor2015united}, technologies that can support smarter and more efficient irrigation practices are expected to have a large impact on global water security.  

Cosmic Ray Neutron Sensing (CRNS) is a relatively new technique that could provide actionable soil moisture information for irrigation scheduling, since CRNS can be deployed above ground to monitor large sites completely non-invasively. Described in \cite{zreda2008measuring}, the technique relies on the use of CRNS probes placed typically \unit{2}{m} above a site to continuously measure the rate of background cosmic ray neutrons back-scattered from surrounding soil. Cosmic ray induced neutrons are strongly moderated by hydrogen, therefore, there exists an inverse correlation between neutron counting rate in the epithermal region (\unit{1}{eV} to \unit{1}{MeV} kinetic energy) and local soil moisture. 
Typically for CRNS campaigns, Helium-3 neutron probes sensitive to thermal neutrons are used, with HDPE moderator shields whose thickness is optimised to  to shift the detectors' dominant sensitivity into the epithermal region of interest \cite{kohli2018response}. In the past 10 years the CRNS technique has become a well-established monitoring method in the hydrological community that provides an effective way to measure the average soil Volumetric Water Content (VWC) over large areas \cite{andreasen2017status}. Recent Monte-Carlo simulations show that the effective radial sensing footprint of a single CRNS probe is in the  region of 120-230~m depending on local conditions \cite{kohli2015footprint}. The COSMOS-US \cite{zreda2012cosmos} and COSMOS-UK \cite{evans2016soil} networks both demonstrate the suitability of the technique for long term monitoring of hydrological conditions, and with more rigorous Monte-Carlo simulations of neutron transport being developed, the precision of the technique has increased to allow for deployment of CRNS probes in more complex applications such as: deployment in built up areas \cite{schron2018intercomparison}, monitoring of canopy biomass \cite{andreasen2017examining}, and scheduling of drip irrigation systems \cite{li2019can}.

Adoption of managed irrigation practices is one way for countries to simultaneously reduce stress on their water resources, increase average yield climate resilience into their water resource supply chain and better manage the resource for expected increases in urban water requirements \cite{zou2012water}. In \cite{ricciardi2018much} it is estimated that 50-70\% of calories consumed globally come from smallholder farms 5 hA or less in size. Unfortunately, given the cost of water usage relative to other growing inputs (e.g. fertilizer, pesticides), adoption of smart irrigation systems in small holder farms is still low, as in many cases the large initial investment required to monitor soil moisture over many hectares and schedule irrigation is significantly higher than the cost savings of the water \cite{schaible2012water}. 

Whilst CRNS moisture monitoring is a powerful technique when  compared to other methods that measure soil moisture over much smaller footprints, the high initial investment required for a CRNS probe is likely to limit widespread adoption of the technique by smallholder farmers. In most studies shown to date, CRNS systems have used Helium-3 detectors due to their large neutron absorption cross-section. The precision of any VWC estimate is a function of the detector's maximum cosmic ray neutron counting rate. As discussed in \cite{weimar2020large}, the minimum integration time required to measure the volumetric soil moisture at the few-percent level can be reduced either by increasing the detection efficiency, $e$, in the signal region of interest (\unit{1}{eV} to \unit{1}{MeV}), or increasing the effective detector volume, $v$. The ideal optimisation metric, $M$, is therefore not purely a function of the overall detector efficiency, but the combination $M$=$ev/c$ where $c$ is the cost per effective volume of the system. Whilst Helium-3 provides one of the highest thermal neutron detection efficiencies available on the market, its high demand means it is not necessarily the optimal detector choice for all applications \cite{kouzes2015progress}. Several groups have already started looking at applying alternatives to Helium-3 detectors developed in the nuclear sector in a hydrological monitoring context. Recently, Boron-lined gaseous detector systems were shown to provide counting rates significantly higher than traditional Helium-3 systems, with a modular design potentially providing additional neutron spectral information on the incoming neutron flux. Moving away from gas based systems, \lifzns~based scintillator detectors are an alternative solution that can be easily scaled to build large area neutron detectors that are optimized for low counting rate applications \cite{barton1991novel}. In these systems Lithium-6 acts as a neutron capture agent, which is mixed with a high sensitivity Zinc-Sulphide scintillator to allow detection of the neutron capture products. Already several companies offer large area \lifzns~ neutron detectors for portal monitor applications \cite{scintacor,symetrica}, an application which shares many of the requirements of the CRNS technique; namely the deployment of robust, temperature stable systems in outside environments. Recent developments of a hybrid EJ-426/EJ-420 (\lifzns) and EJ-299-33 (pulse-shape discrimination plastic scintillator) based detector system found it was possible to construct a mixed field radiation sensitive to both gammas, muons, and thermal and fast neutrons. This system was found to have  neutron detection efficiencies of approximately 55\% that of a typical Helium-3 based CRS1000 neutron detector. This presents a promising first step in developing scintillator based alternatives to Helium-3 in the CRNS agriculture sector, as there exists extensive prior work in the development and optimisation of \lifzns~scintillators \cite{yehuda2018optimization,marsden2013large}, their associated trigger systems \cite{stoykov2016trigger,cowles2018development}, and correlations between muon and neutron cosmic ray intensities \cite{de2016temperature,maghrabi2012kacst,abdullrahman2020first}.

This paper assesses the feasibility of constructing low-cost neutron sensors, intended for soil moisture monitoring, out of scintillating foils made from \lifzns~or \bnzns. In contrast to the detector shown in \cite{stevanato2019novel}, the detectors are constructed from \lifzns~foils sensitive to only the thermal component of the cosmic ray flux, and therefore their energy sensitivity is closer to that of a typical Helium-3 tube. By focusing on only the thermal neutron region they are expected to have significantly lower cost and power requirements compared to mixed neutron/gamma radiation detectors since fast digitisers with built-in Pulse-Shape-Discrimination (PSD) are not required for optimal particle identification. Furthermore, replacing the Li$^6$F with un-enriched BN to assess the improvement in cost versus efficiency when using readily available detector materials has been investigated. The paper is structured as follows; Section \ref{sec:detectordesign} describes the sensitivity of the neutron sensitive foils and the main detector design, Section \ref{sec:simulatedefficiency} investigates the expected detector efficiency based on GEANT4 simulations, Section \ref{sec:lightcollection} evaluates the effect that different light guide designs have on simulated detector sensitivity, Section \ref{sec:detectorconstruction} gives an overview of the detector construction and testing and Section \ref{sec:conclusion} summarises the results and future potential applications of these detectors.

\section{Detector Design}
\label{sec:detectordesign}
\begin{figure} 
    \centering
    \includegraphics[width=0.9\textwidth, clip, trim= 0 280 0 0]{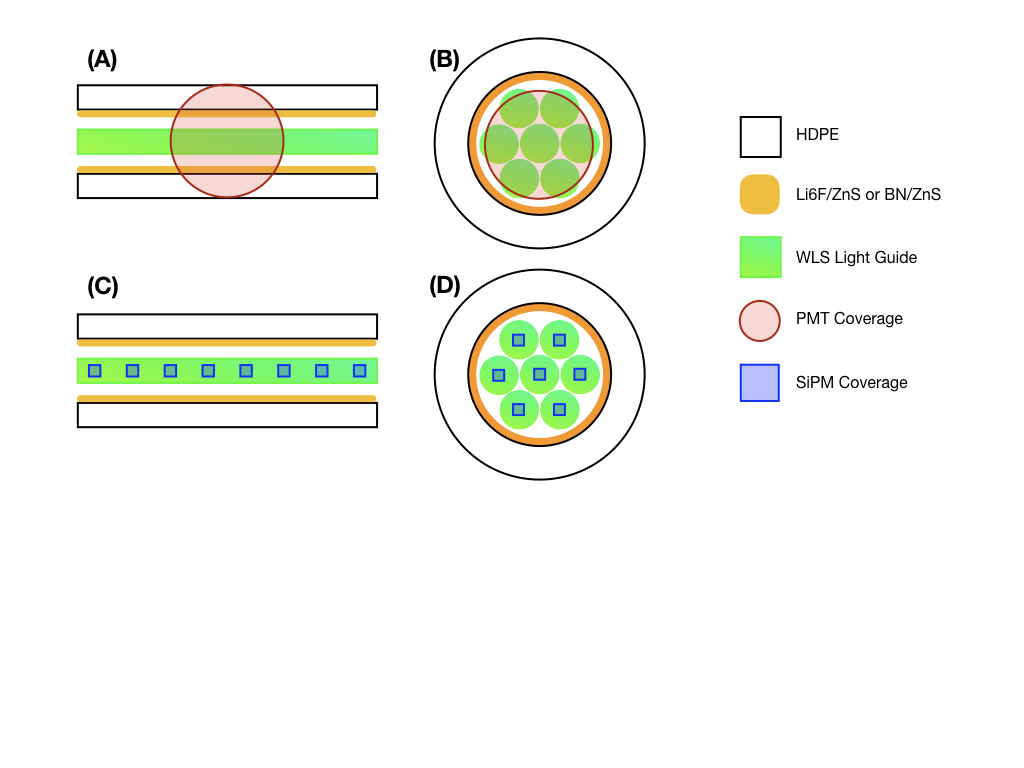}
    \caption{Comparison of possible "planar" (A and C) and "cylindrical" (B and D) detector geometries when constructing a scintillator-based neutron detector.}
    \label{fig:geometrydiagram}
\end{figure}

In neutron portal monitor applications, planar wavelength shifting neutron detector configurations have typically been used to build large area systems \cite{symetrica}. The use of internal reflection inside  planar light guides allows large sheets of \lifzns~to be optically coupled to a Photo-Multiplier Tube (PMT) at one end with a much smaller photo-cathode area as shown in Figure \ref{fig:geometrydiagram}. This helps to reduce the overall cost per unit area of the system. Wavelength-shifting plastic helps to improve overall system efficiency even in the case where the emission wavelength is at a lower PMT quantum efficiency, since the likelihood of internal reflection is significantly higher if the light is emitted isotropically from within the light guide itself. Since neutron discrimination relies on detection of a train of a few photon pulses from the long ZnS decay time (of the order of \unit{10}{$\mu$s}), good optical transmission is needed when constructing large planar detectors. The cost per unit area relative to traditional Helium-3 detectors is therefore dominated by the best achievable absorption length and surface quality of the light guides.

\begin{figure} [h]
    \centering
    \includegraphics[width=0.8\textwidth, clip, trim= 0 30 0 0]{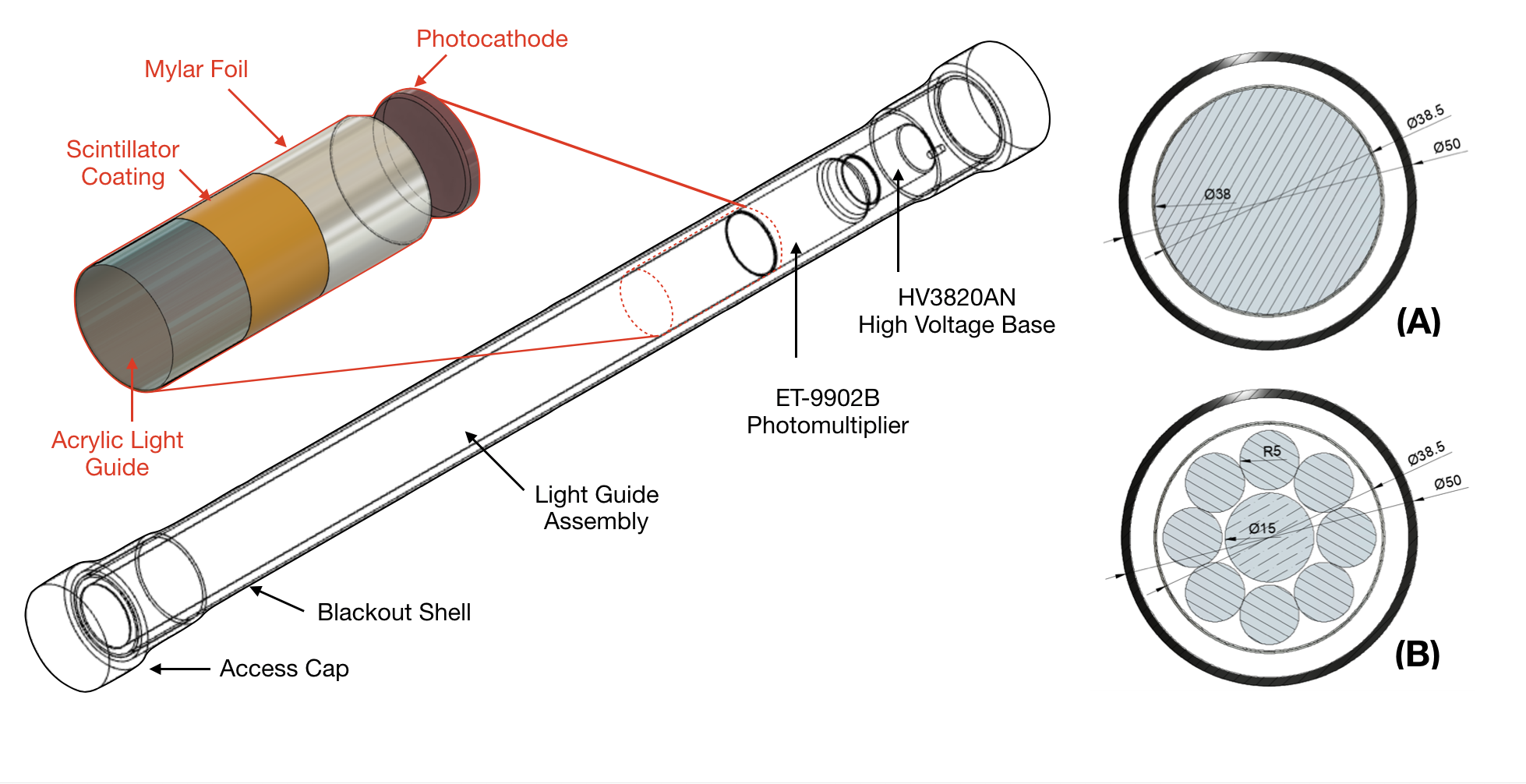}
    \caption{(left) Main cylindrical detector components. (right) The Cross-section of the light guide assembly for a 38~mm cylinder (A) and segmented "nine" rod collection geometry (B). }
    \label{fig:detectordiagram}
\end{figure}

In this paper, instead of planar detectors, the efficiency of an alternative cylindrical neutron detector design, more closely resembling a traditional Helium-3 tube as shown in Figure \ref{fig:detectordiagram} is considered. Adopting a long cylindrical geometry in this fashion is expected to reduce the overall cost of a detection system when produced in volume. By constructing the light guides from readily available  acrylic rods instead of scintillator such as EJ299-33, the detectors are expected to be predominantly sensitive to thermal neutron capture events which leads to long scintillator decay times in the ZnS. The reduction in sensitivity to gamma/muon backgrounds means thermal neutrons can easily be discriminated using the time over threshold in a low cost discriminator circuit. Furthermore, the use of cylindrical light guide assemblies allows extruded plastic to be used in place of cast PVT based EJ280, reducing overall cost.

The alternative detector design considered (shown in Figure \ref{fig:detectordiagram}), consists of a central light guide assembly, wrapped in a scintillating thermal neutron foil, and coupled to an ET-Enterprises ET-9902B 38~mm diameter PMT at one end. Whilst a ET-9902B PMT has been selected for this work, the advantage of this cylindrical design is that it can be easily adapted to smaller or larger PMTs without significant internal redesign, and space-efficient outer enclosures can be constructed from light-tight tubing at low cost. Different light guide structures have been considered to assess the improvement wavelength shifting materials can make to a cylindrical detector sensitivity. In addition, the question of whether segmentation of the internal light guide can further improve the detector light yield for neutron capture events has been investigated. In \cite{marsden2013large} it was shown that fluorescent acrylic rods arranged in a grid and individually wrapped in thermal neutron scintillator could increase neutron capture light yield by up to 50\% when compared to a planar light guide configuration. In this study light guides that are wrapped as one bundle by a single scintillating foil are considered, as this allows a direct comparison of the performance of the cylindrical and segmented light guide geometries. A 38~mm diameter solid cylinder, and collections of either seven or nine smaller rods with radii chosen to optimise the coupling to a 38mm PMT were considered as possible configurations.

\section{Simulated Capture Efficiency}
\label{sec:simulatedefficiency}

Using GEANT4, the efficiency of each of the proposed thermal neutron scintillator geometries in the energy region of interest for soil moisture monitoring is evaluated. Each detector geometry has been constructed from a $100~\mu\mbox{m}$ thick scintillator tube with outer diameter 38.2~mm. Scintillator material compositions are based on mixing proportions given by weight in \cite{barton1991novel,marsden2013large}.  These tubes are placed around each of the simulated 50~cm long cylindrical light guide geometries. The  detectors are surrounded by a moderator shield of variable thickness constructed from high density polyethylene (HDPE). In addition, a planar geometry is considered which uses the same volume of scintillator as in the cylindrical case, but instead is spread into two 5.96~cm x 50~cm sheets placed either side of a 10~mm thick PVT light guide and stacked between two variable thickness layers of HDPE sheets. HDPE is assumed to have the density and composition  of GEANT4's  NIST  material  managers G4\_POLYETHYLENE  definition, however lower density polyethylene (LDPE) is also considered, with a density of only 0.55~g/cm$^3$. LDPE approximates the construction of a cylindrical moderator shield constructed out of readily available HDPE beads with a packing fraction of 58.5\%. 


Neutrons are generated isotropically with a random direction from the surface of a $1\mbox{ m radius}\times2\mbox{ m}$ height cylinder surrounding the simulated detector. Random directions are used to include events in which a neutron may enter the detector at a shallow angle on the outer edge of the moderator shield, but then undergo a random scatterings and end up in the inner detector. A random starting kinetic energy between $0.01$~eV and $10^7$~eV is chosen. The propagation of these neutrons through the detector is simulated using the QGSP-BERT-HP physics model, and the total energy deposited by each neutron inside the scintillator is recorded. A neutron is assumed to have been detected if approximately 4~MeV (2~MeV) of energy is deposited in a \lifzns~(\bnzns) layer. This neglects all effects relating to scintillator light transmission and therefore gives only an estimated capture efficiency based on the geometry of the scintillator layers and surrounding moderator. Due to the randomly chosen direction, a large proportion of neutrons miss the detector entirely. To account for this, the total neutrons detected for each geometry are divided by that predicted in a secondary simulation of a hypothetical perfect cylindrical detector (20~cm radius x 100~cm length) placed in the centre of the geometry. All efficiencies shown are  relative to this hypothetical "true" detection system. 

\comment{https://tsapps.nist.gov/publication/get_pdf.cfm?pub_id=923899}

\begin{figure}[h]
    \centering
    \includegraphics[width=0.48\textwidth]{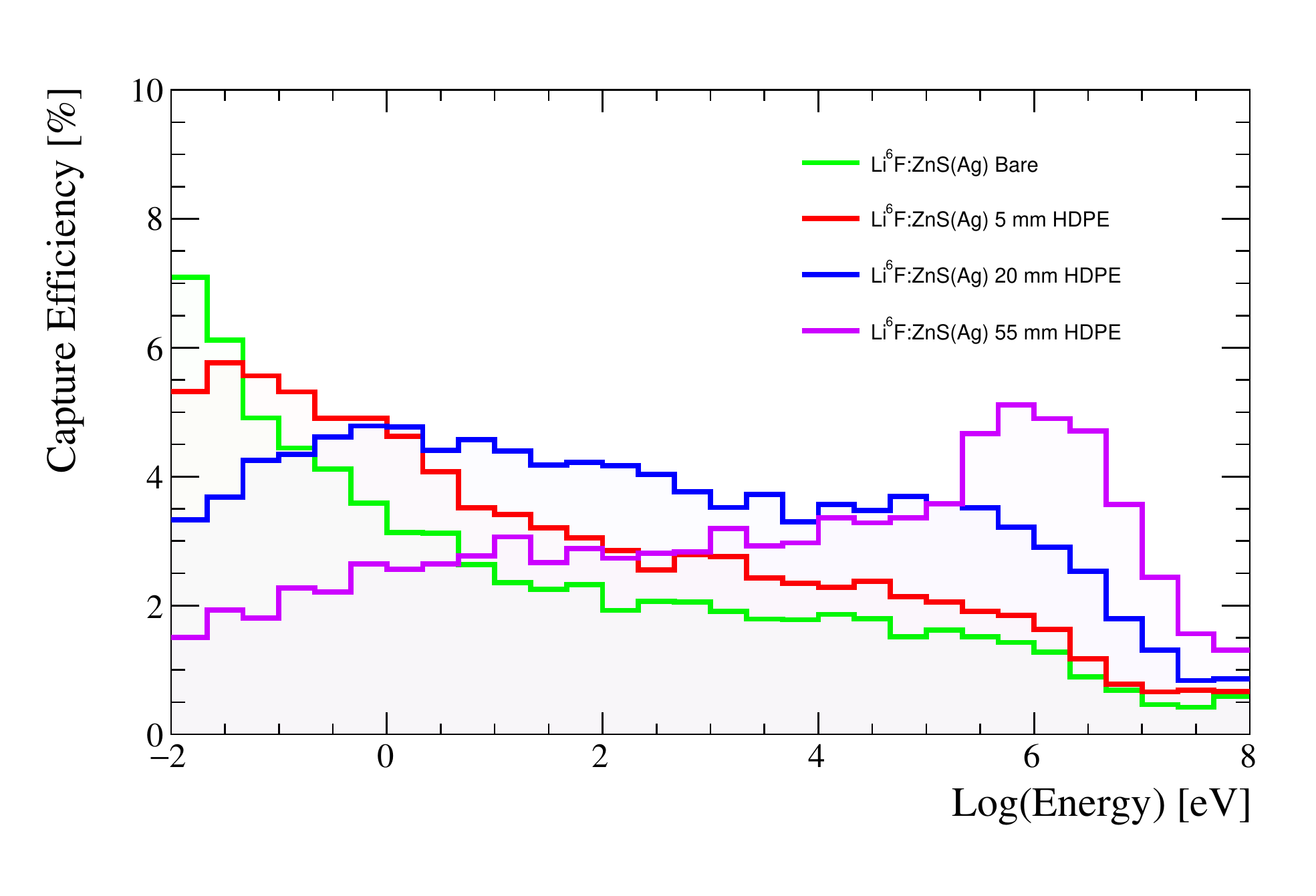}
    \includegraphics[width=0.48\textwidth]{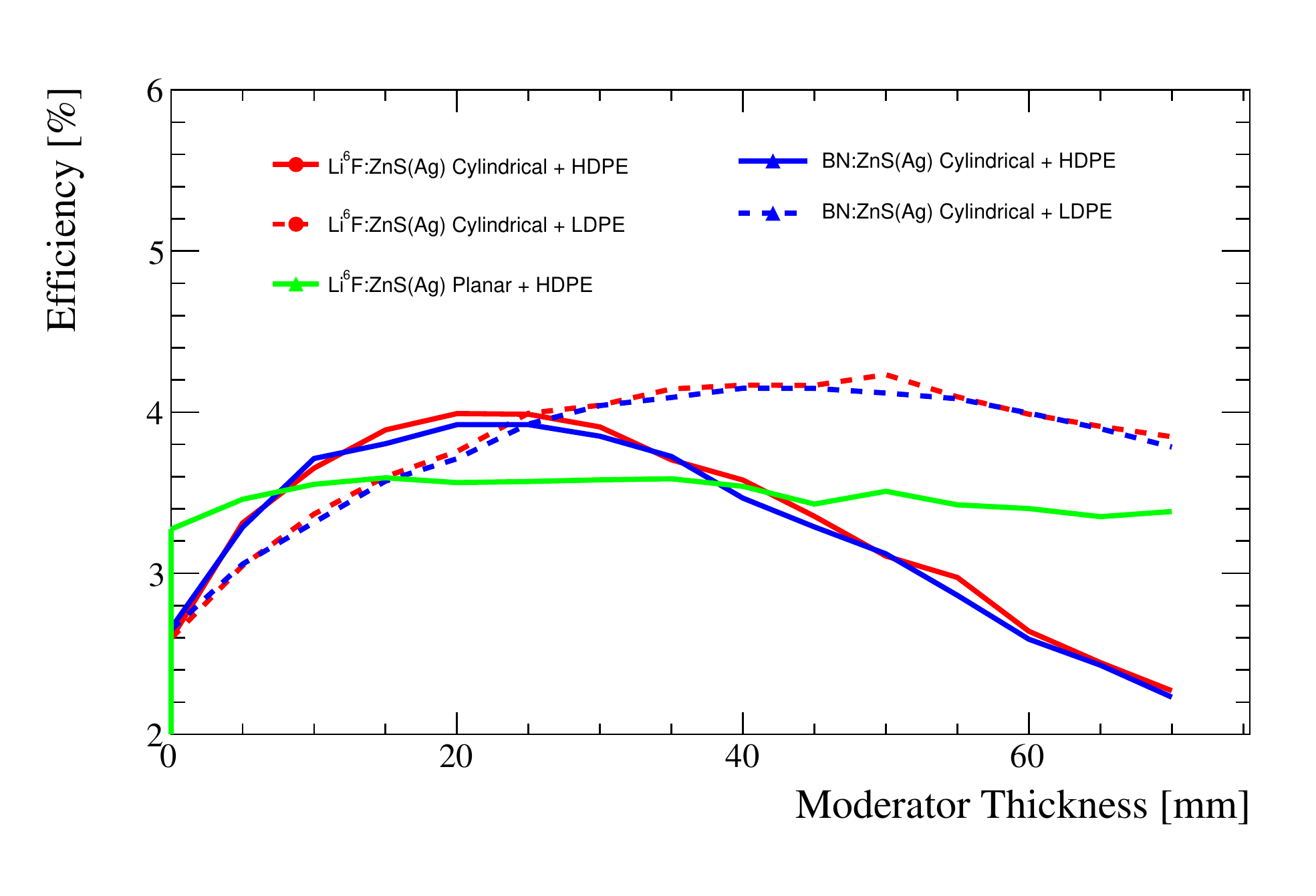}
    \caption{Capture efficiency of the different detector geometries considered in this work. (left) Capture efficiency of a cylindrical \lifzns~detector as a function of starting  neutron kinetic with varying thicknesses of HDPE moderator. (Right) Average  efficiency in the region of interest as a function of HDPE/LDPE thickness for the detector geometries considered in this work. }
    \label{fig:modefficiency}
\end{figure}

As shown in Figure \ref{fig:modefficiency}, the efficiency of a bare scintillator detector system is maximal at thermal neutron energies as this is where the \lifzns~(\bnzns) capture cross-section is highest for both planar and cylindrical geometries. 
The GEANT4 simulations show an increased efficiency tail out to $10^6$~eV due to a small number of neutrons losing energy due to interactions in the light guides before being captured. No significant differences were observed due to the slight reduction in overall detector mass when using segmented internal light guides for the cylindrical detectors, therefore comparison of the results is neglected here.

 To have an optimal sensitivity in the region of interest for soil moisture monitoring both cylindrical and planar light guide geometries require additional external neutron moderator. Simulations of different moderator thicknesses suggest an optimal HDPE thickness of 30~mm for the planar systems and 20-25~mm for the cylindrical systems. This is consistent with similar studies performed for Helium-3 based neutron detectors in \cite{kohli2018response}. In addition, it is observed that LDPE moderator shields are also suitable for a cylindrical detector given larger thicknesses of 45~mm can be used. However, doing so also results in a slightly higher background thermal detection efficiency. Whilst the peak signal efficiency was found using 45~mm LDPE, comparable efficiencies to the HDPE case were found using as little as 30~mm of LDPE. Since the thermal neutron component is from evaporation neutrons,  maximal signal to noise ratio is given when the capture efficiency of incoming neutrons is maximised in the 1-6 MeV region and minimised in the thermal 0.01~eV-1~eV region. Thermal neutron-absorbing cadmium has been suggested as an way to suppress the thermal counts and improve signal to noise \cite{desilets2010nature}, however this is neglected in the present work. The advantage of using lower density HDPE is that a moderator shield can be constructed entirely from easily available 3~mm HDPE beads, allowing rapid modification of the moderator choice for a specific application without requiring separately cast outer HDPE shields for each application. For future iterations of this detector it is planned to investigate the combination of HDPE beads and boron lined outer shells to produce a low cost and non-toxic optimal moderator solution. 
 
When considering the optimal moderator choice both cylindrical and detector geometries show a comparable efficiency when using the same volume of scintillator in each one. Whilst the total surface area of the front face of the planar detector is over 50\% larger than the cylindrical detector, the  efficiency averaged over all angles is similar due to the thin cross-section of the planar detector on one side. Multiple planar detectors placed at varying angles in a field could take advantage of this angular effect to extract variations in the neutron rate as a function of direction, however doing so with a high precision would likely require much larger detectors and a higher resolution soil sampling campaign to calibrate for their varying footprints in the field.  In agreement with studies shown in \cite{marsden2013large}, competitive efficiencies are obtained for the un-enriched \bnzns~detector simulations systems  when compared to \lifzns, confirming the suitability of the material in production of low-cost neutron detectors provided scintillator detectors can be manufactured with high enough sensitivity to account for the lower scintillation light yield of \bnzns. 

Whilst promising, the presented efficiency comparisons for planar and cylindrical detectors assume a 100\% triggering efficiency once a neutron captures in the foils. In the next section the optimal light collection system for a cylindrical detector is evaluated in order to understand the maximum length of detector that could be constructed.

\section{Light Guide Simulations}
\label{sec:lightcollection}
\begin{figure}
    \centering
    \includegraphics[width=0.48\textwidth]{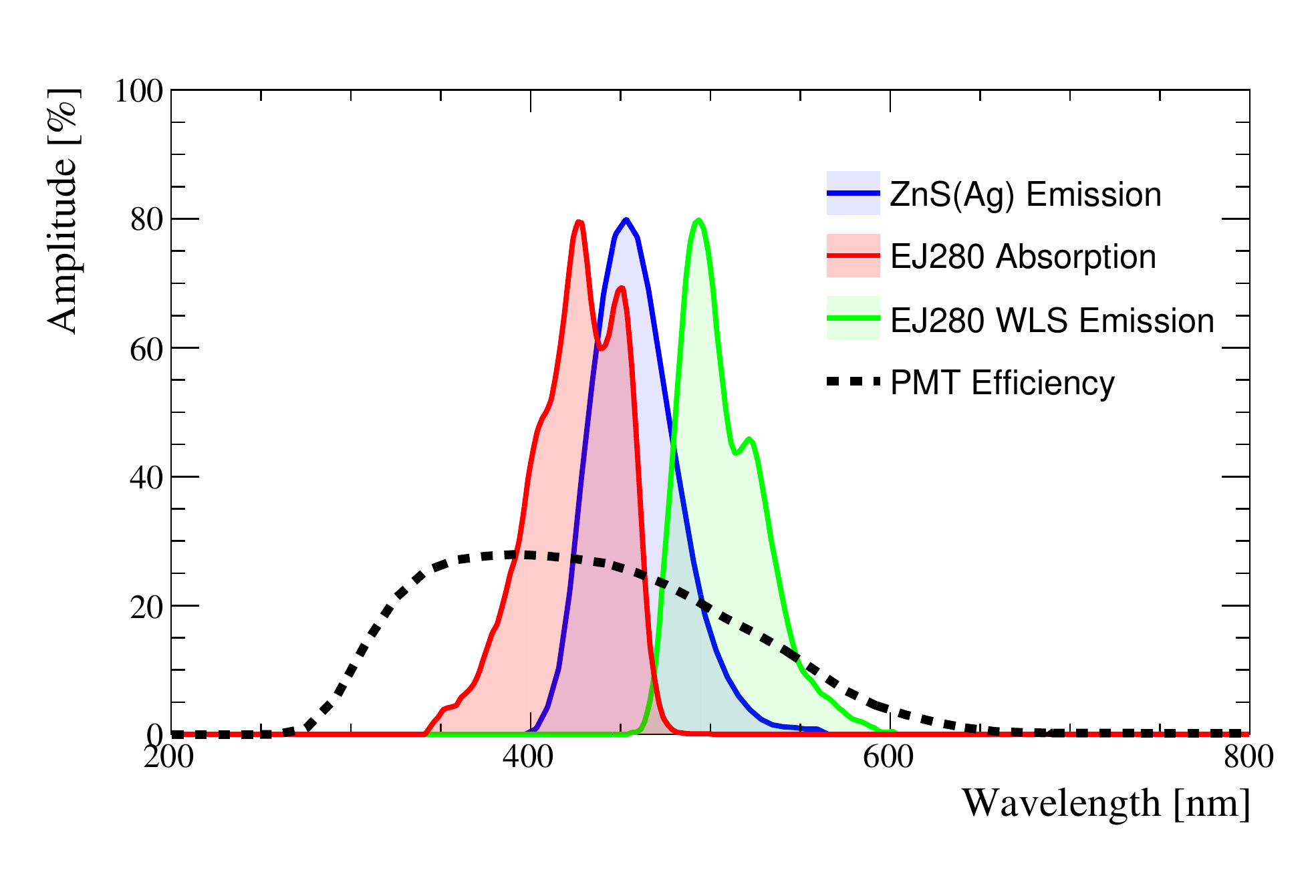}
    \includegraphics[width=0.48\textwidth]{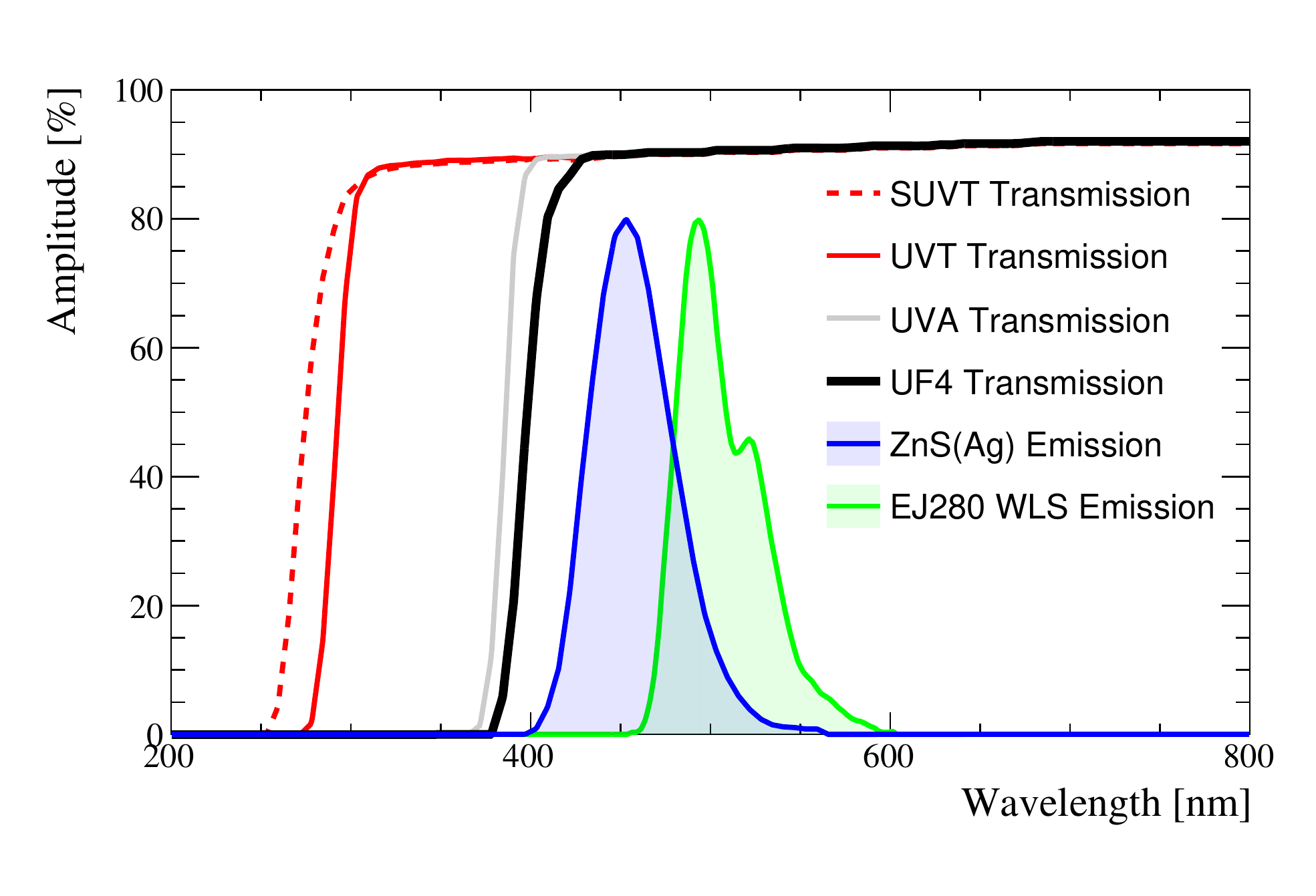}
    \caption{(left) Emission Spectrum of EJ280, ZnS and PMT efficiency. (right) transmission properties of materials under consideration in this work.}
    \label{fig:fluorspectra}
\end{figure}

To understand the effect light guide geometry has on overall light collection efficiency, simulations of scintillator emission for \lifzns~and 
\bnzns~composites and a range of light guide materials using GEANT4's optical tracking capability were performed. A modification of the QGSP\_BERT\_HP physics list was made to include Cerenkov and scintillation light production and wavelength shifting. Optical boundaries were added to the cylindrical detector geometries discussed in the previous section using the RealSurface dynamic look-up-table, modelling the scintillator surface as a ground-to-air border, and the light guide surface as a polished-to-air border. Absorption  and  emission curves for each material were included as look-up tables digitized from supplier sources (Figure \ref{fig:fluorspectra}). Full details of material properties included in the GEANT4 simulation are shown in Table \ref{tab:geant4light}.

\begin{figure}
    \centering
    \includegraphics[height=0.37\textwidth, clip, trim=0 450 690 20]{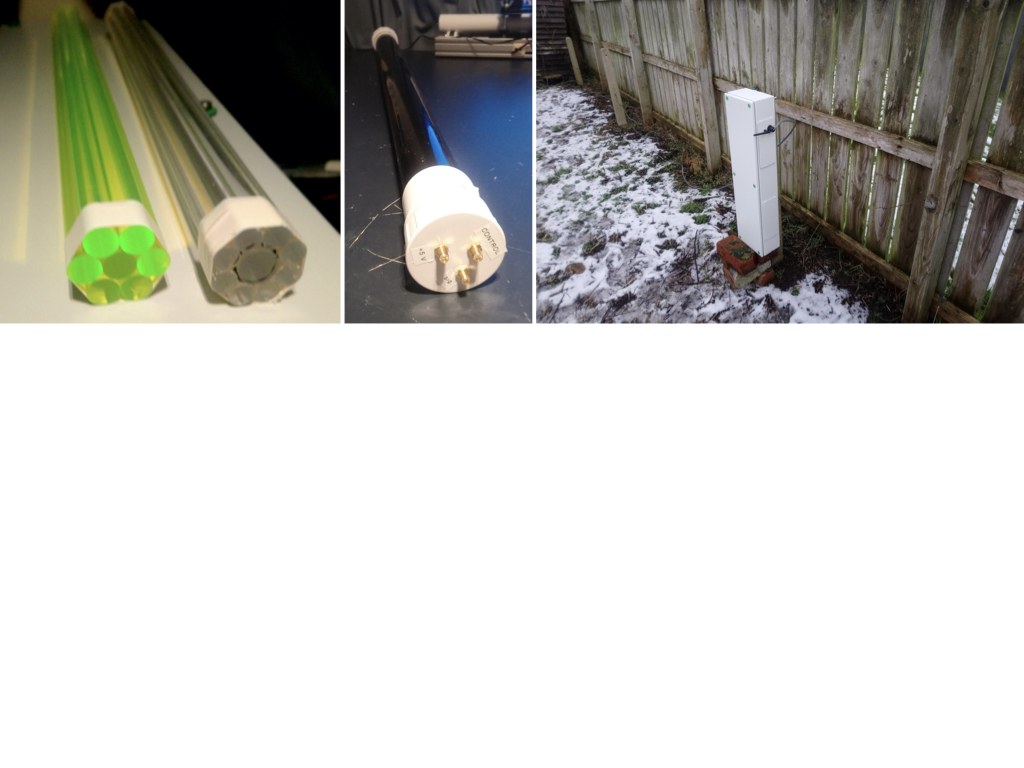}
    \includegraphics[height=0.37\textwidth]{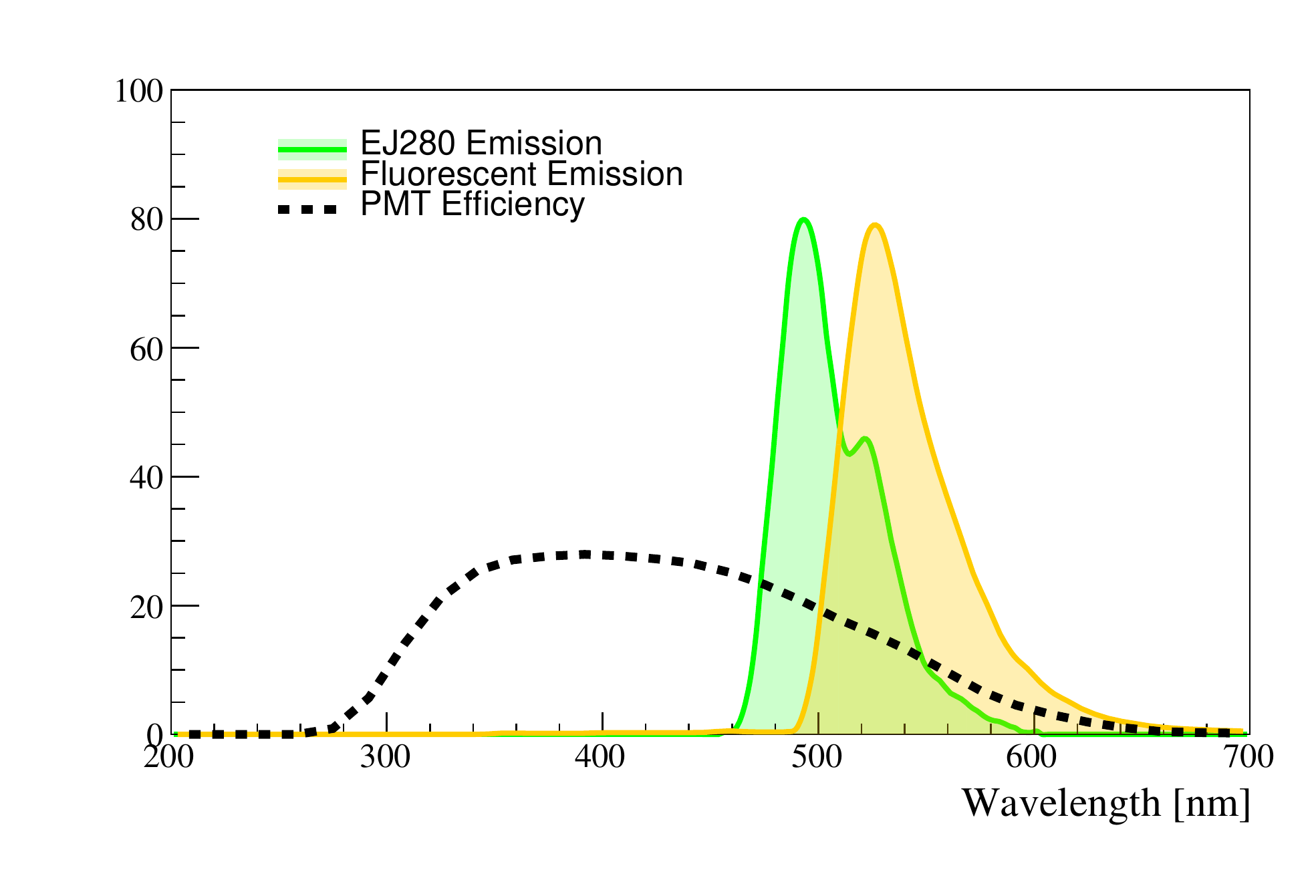}
    \caption{(left) Two different light guide geometries considered and difference between fluorescent and undoped acrylic. (right) Difference in emission spectrum for commercially available fluorescent rods and EJ280.}
    \label{fig:shortspectra}
\end{figure}

\begin{table}
    \centering
    \begin{tabular}{| c | c | c | c | c | c | }
    \hline
    Material & R Index & Max. Absorption Length & Emission Yield & Peak Emission \\
    \hline
        \lifzns &  2.36 & 0.15~mm & 47000/MeV & 440~nm \\
        \bnzns & 2.36 & 0.15~mm & 47000/MeV & 440~nm \\
        EJ280 & 1.58 & 3.5~m & - & 490~nm \\
        Clear PMMA & 1.49 & 3.1~m & - & - \\
        Flourescent PMMA & 1.49 & 57.56~m & - & 550~nm \\
        \hline
    \end{tabular}
    \caption{ Geant4 material definitions used in this work. }
    \label{tab:geant4light}
\end{table}
As shown in Figure \ref{fig:fluorspectra}, ZnS scintillator emits in the 400~nm to 430~nm region. The 9902B photo-multiplier tube selected for this work has sensitive region which peaks in the 430~nm region and so is well-matched for direct detection of the ZnS scintillator, however it has a width which extends out to 300-600~nm at half maximum, covering the wavelength shifting emission spectrum of EJ280. The transmission efficiency of different un-doped clear acrylics can have very different transmission curves in the VHUV region, however these differences are reduced at wavelengths greater than 400~nm so are not expected to have a significant effect on detector performance. In addition to EJ280 and the available clear PMMA material, an additional material referred to simply as "Fluorsecent" which is based on commercially available fluorescent rods was also considered. In \cite{marsden2013large}, it was shown that off-the-shelf fluorescent rods performed substantially better than dip-coated wavelength shifting rods when trying to produce a detector 50-60~cm long. To verify this, a small quantity of  fluorescent rods was acquired from a local supplier. Using a Flame-NIR-VIS spectrometer and 400~nm UV LED to test the emission spectrum of these rods, it was observed that the spectrum is shifted slightly towards the yellow, peaking at approximately 550nm as shown in Figure \ref{fig:shortspectra}. Similar results  were seen in \cite{marsden2013large} which were hypothesised to be a result of a larger proportion of fluorescence doping  compared to  EJ280 leading to a higher probability of re-absorption of emitted photons in the 450-500~nm  region. The absorption length of these fluorescent rods when considering the relative change in overall light transmission for all wavelengths for different distances was determined to be 57.56 $\pm$ 5.68 cm. 

Simulations of light yield were performed by generating a logarithmic energy distribution of neutrons in the same way as described in Section \ref{sec:simulatedefficiency}, and directing them at the cylindrical detector geometries, now with a full description of scintillator emission and optical tracking throughout the detector included. For each event the total number of photons detected at the PMT placed at one end of the light guide was recorded alongside the location of the true energy deposit position inside the scintillator layer. The PMT quantum efficiency distribution was accounted for by randomly throwing away photons at the simulation analysis stage based on the arriving photon wavelength. The number of remaining photons were then used to build a map of the reduction in light collected as a function of neutron capture distance along the light guide. Since the scintillator is assumed to be a homogeneous solution of \lifzns~(\bnzns) these maps do not include any variations in sensitivity due to microscopic variations in the scintillator (for example due to choice of Zn or Li$^6$F/BN grain size, or variations in scintillator thickness across the detector). 
\begin{figure}
    \centering
    \includegraphics[width=0.45\textwidth]{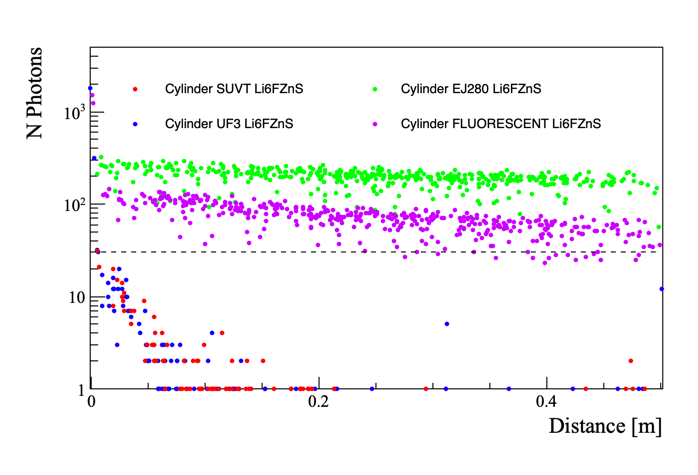}
     \includegraphics[width=0.45\textwidth]{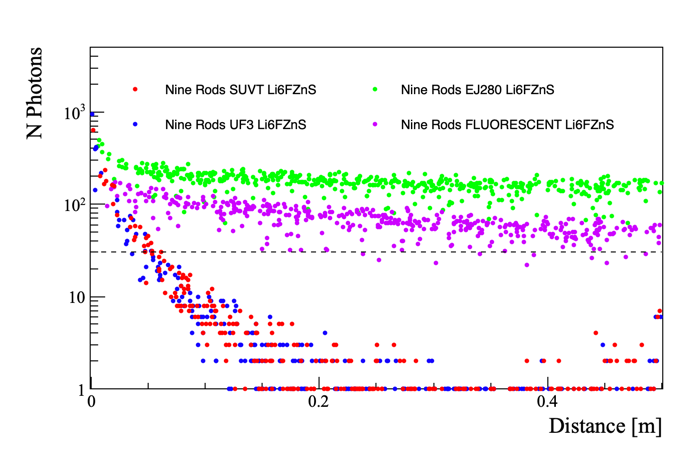}
     \includegraphics[width=0.45\textwidth]{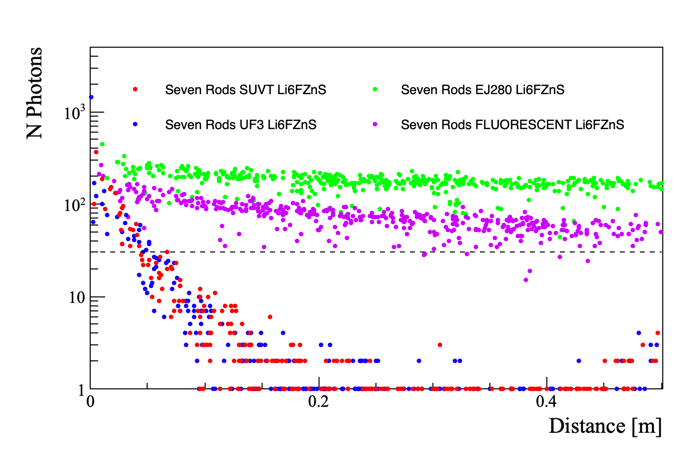}
     \includegraphics[width=0.45\textwidth]{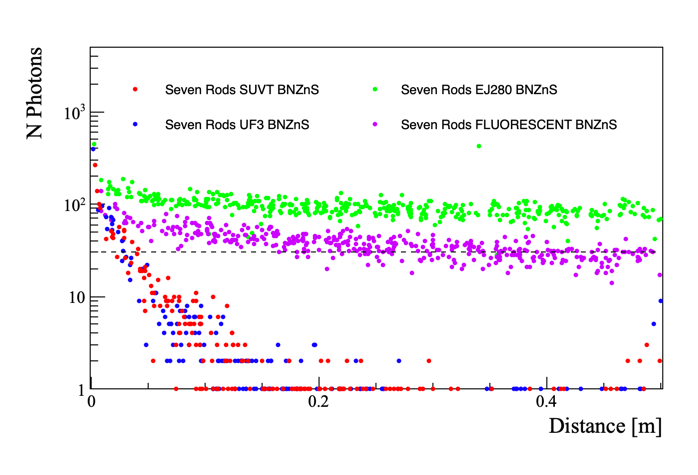}
    \caption{Number of emitted photons as a function of distance for different light guide geometries and construction materials. Each point corresponds to a single simulated event. (Top-left) Cylindrical geometries with Li6FZnS scintillator. (Top-right) Nine rod geometries with Li6FZnS scintillator. (Bottom-left) Seven rod geometries with Li6FZnS scintillator. (Bottom-right) Seven rod geometries with BNZnS scintillator. The dashed line gives a guide for a 30 photon cut-off.}
    \label{fig:foilefficiency}
\end{figure}

As expected, a reduction in the number of photons observed with increasing distance is observed. Most notable however is the large excess of photons within 5-10~cm of the PMT face which are a result of photons emitted from the scintillator travelling directly into the PMT without being captured in the light guide. In the case of clear acrylic light guides, capture inside the acrylic is minimal, since any scintillation photons created at an angle suitable for total internal reflection have a high probability of reflecting off the exterior of the acrylic light guide back towards the opaque scintillator coating. As a result all geometries using acrylic SUVT or UF3 materials observed a sharp drop in the total photons within the first 10 cm of the detector. Segmented light guides showed a slightly improved response.

To estimate the maximum recommended detector size a hard cut requiring at least 30 scintillation photons to be detected was implemented. This is a somewhat arbitrary, however it is  based on work in \cite{stoykov2016trigger}, where a multi-coincidence trigger applied to the \lifzns~scintillator emission was shown to have maximal efficiency when requiring approximately 10 consecutive single photon trigger stages. A larger but similar order of magnitude cut of at least 30 single photons is deemed reasonable to provide an approximate guide, given that several other inefficiencies are neglected in this work (light guide roughness, coupling inefficiencies, end mirror optimisation). This cut suggests that when using a clear light guide which does not include any wavelength shifting material, the maximum detector length is likely to be 12~cm, with over 75\% of the scintillator layer producing very weak scintillation events that may be difficult to detect. When using a single cylindrical light guide the maximum length is even further reduced to only 5~cm. For all light guide materials considered no significant difference in efficiency is seen for a "nine" or "seven" rod configuration.

Light guides with an absorption length and emission spectrum matching EJ280 showed a significantly improved light collection efficiency for all geometries that extended to the full 50~cm proposed length of the detector when assuming an absorption length of 3.5~m. Detectors using a segmented wavelength shifting central light guide showed a slightly sharper drop in light yield for neutron captures within the first 10 cm of the detector when compared to a bulk cylindrical light guide. This difference is relatively small however, and is likely to be insignificant when compared to the higher expected cost to obtain large diameter wavelength shifting cylinders. The "Fluorescent" light guide geometry also showed a significantly improved light collection efficiency compared to the clear acrylic case, despite the shift of photons to a region of lower PMT quantum efficiency. As expected, the \bnzns~detector simulations show a light yield approximately 50\% that of \lifzns, highlighting that a maximum detector size of 7~cm is recommended without using wavelength shifting light guides to improve light collection efficiency, whilst detectors approximately 30~cm long may still be appropriate when utilising the low cost fluorescent rods. 

These studies suggest that using \lifzns~foils coupled with collections of small diameter, low cost fluorescent light guides, will still result in neutron captures producing events with roughly 30-100 photons within the first 30~cm of the detector. In the next section the development of a simple portable detector system for these detectors based on this configuration is discussed.

\comment{Light guide efficiency for different UVT properties taken from here : https://www.google.com/imgres?imgurl=https
}

\section{Detector Prototypes}
\label{sec:detectorconstruction}
\begin{figure}
    \centering
    \includegraphics[width=1.0\textwidth, clip, trim=0 220 0 0]{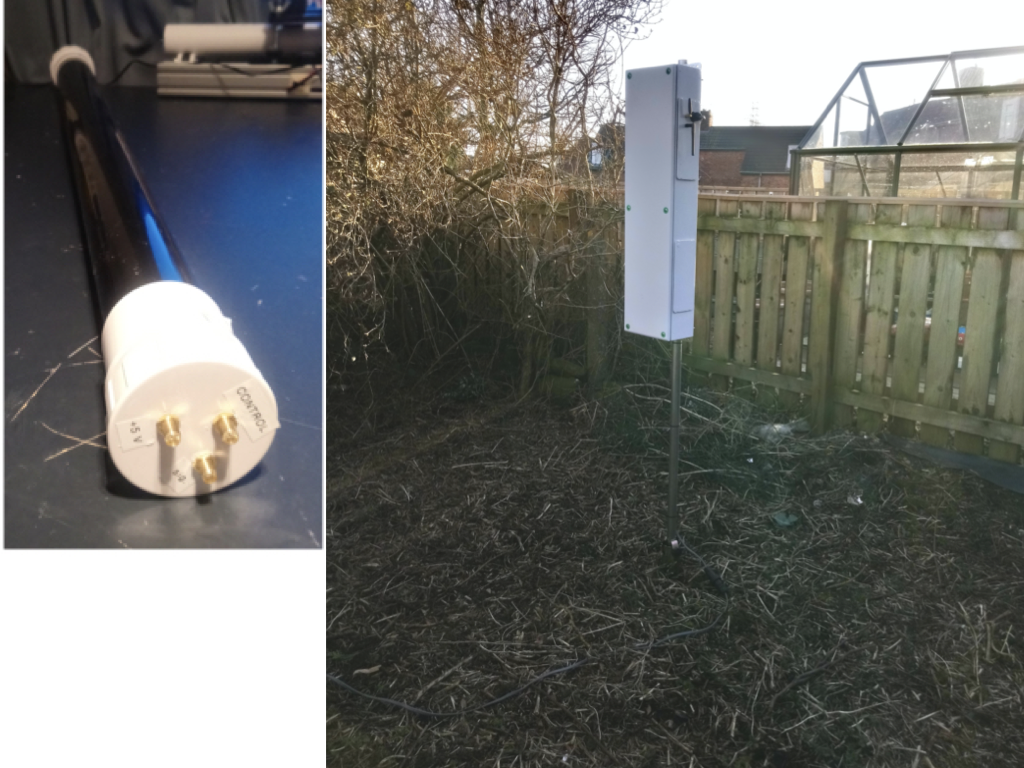}
    \caption{Assembled detector prototype. (left) Bare detector tube before being surrounded in LDPE moderator. (right) Assembled IP55 weatherproof detector system mounted onto a pole. }
    \label{fig:detbuild}
\end{figure}

Based on the simulations in the previous section a detector prototype was constructed from fluorescent "WLS" rods and \lifzns~scintillator foils. This wrapped light guide geometry was coupled to an ET-Enterprises 9902B PMT, and placed inside a light-tight outer detector housing. The assembled internal light guide geometries and final assembled detector tube are shown in Figure \ref{fig:detbuild}. An external moderator shield was constructed from a 100~mm PVC tube and filled with HDPE beads to give an effective moderator thickness of 30~mm following the studies described in Section 2. Pulses from each detector were digitized using a DRS4 1-GHz digitiser connected to a Raspberry Pi single board computer. The High Voltage (HV) bias value of each PMT was calibrated using a digital to analog convertor connected to the HV base to give an equivalent gain for each PMT. The total power consumption of the PMT, digitiser, and single board computer was found to be less than 4W. This integrated system was installed inside an IP55 weatherproof case, resulting in a compact and mobile system ready for outdoor use as shown in Figure \ref{fig:detbuild}.

\begin{figure}
    \centering
    \includegraphics[width=0.49\textwidth]{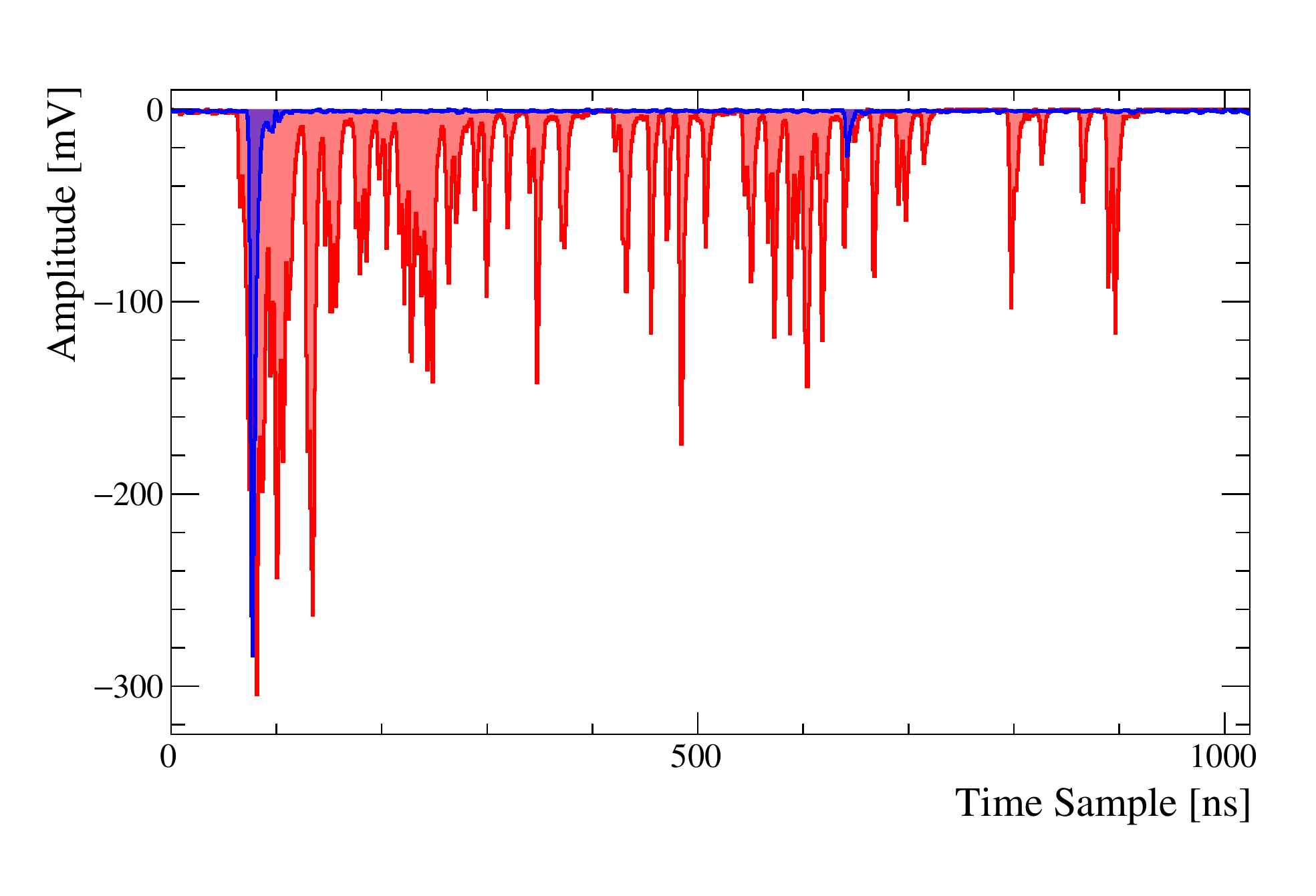}
    \includegraphics[width=0.49\textwidth]{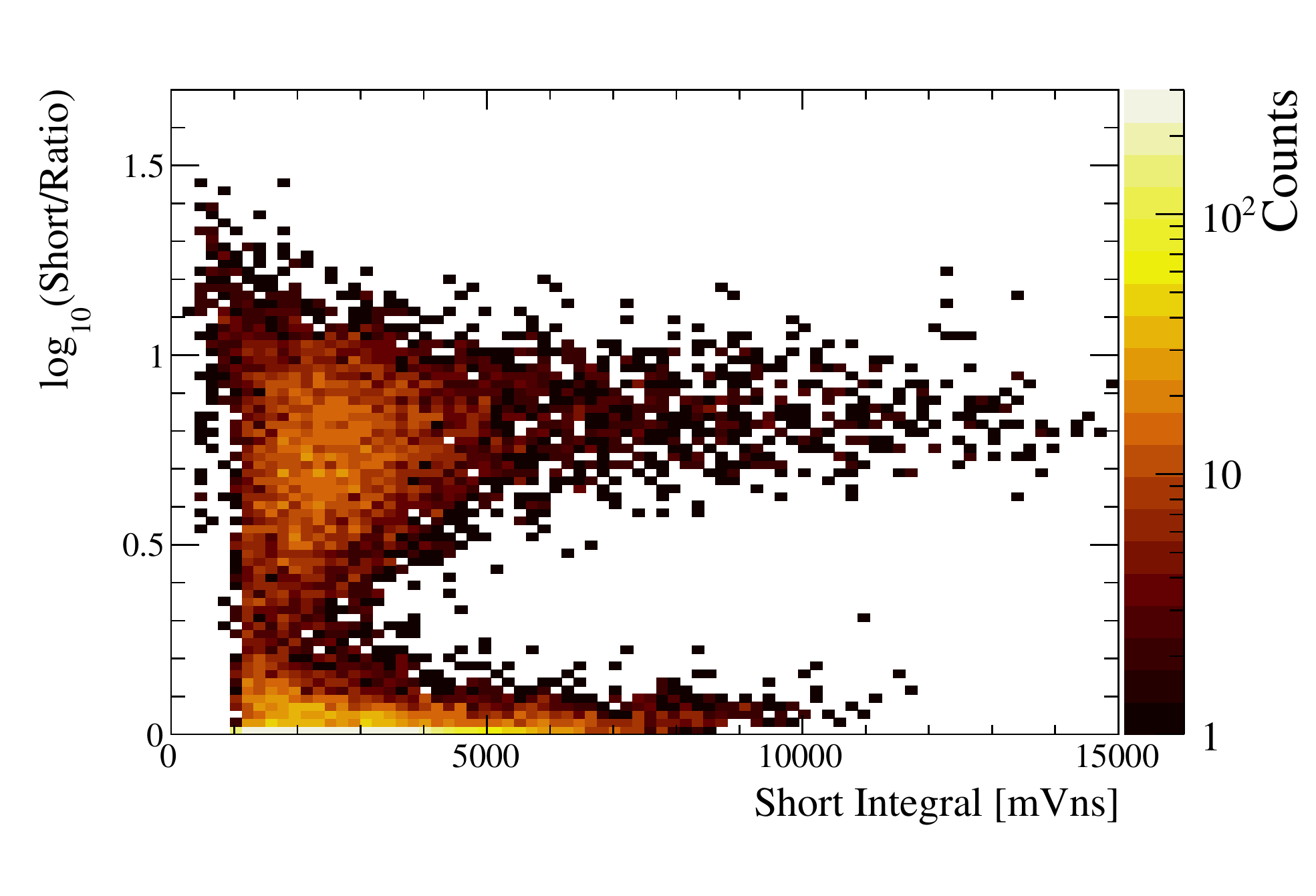}
    \includegraphics[width=0.49\textwidth]{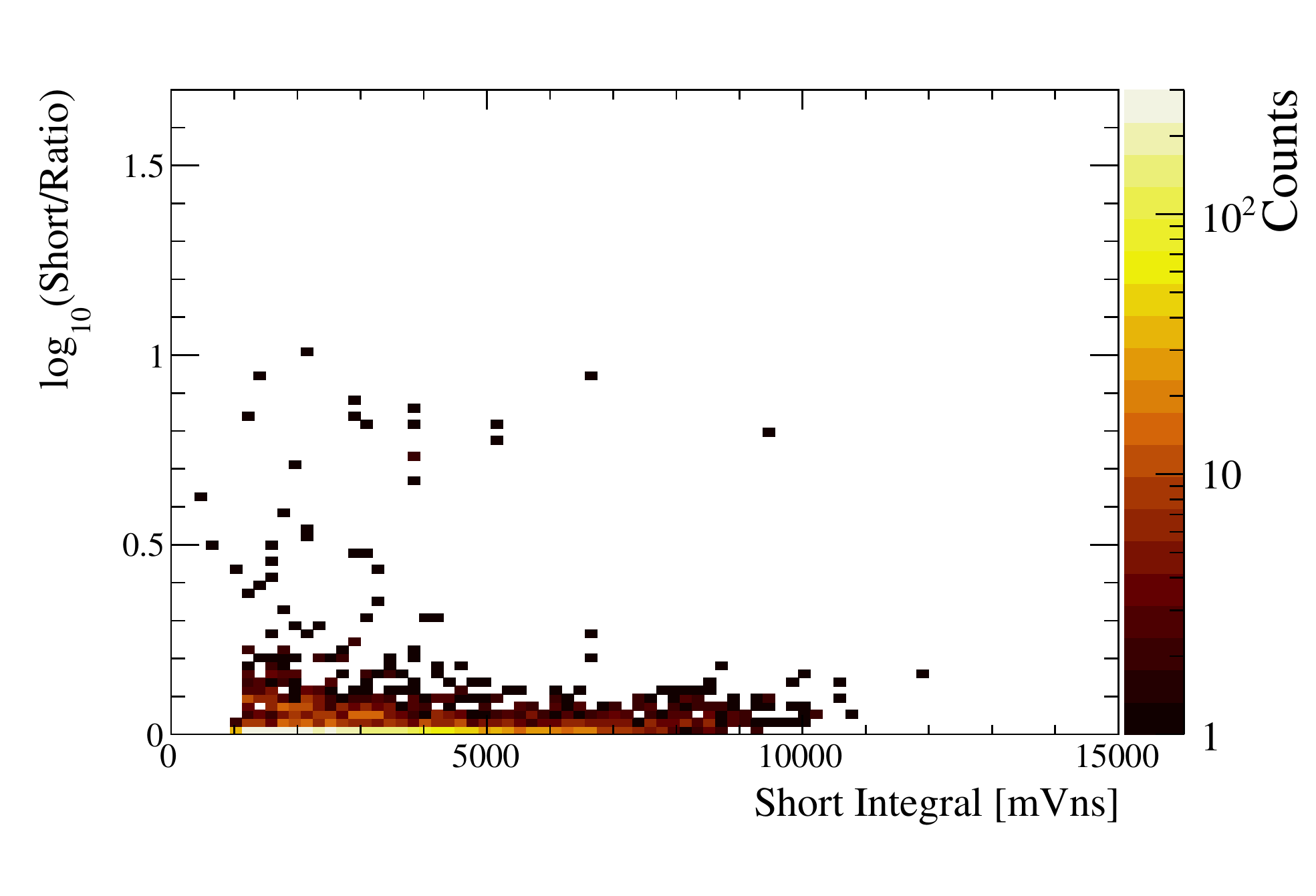}
    \includegraphics[width=0.49\textwidth]{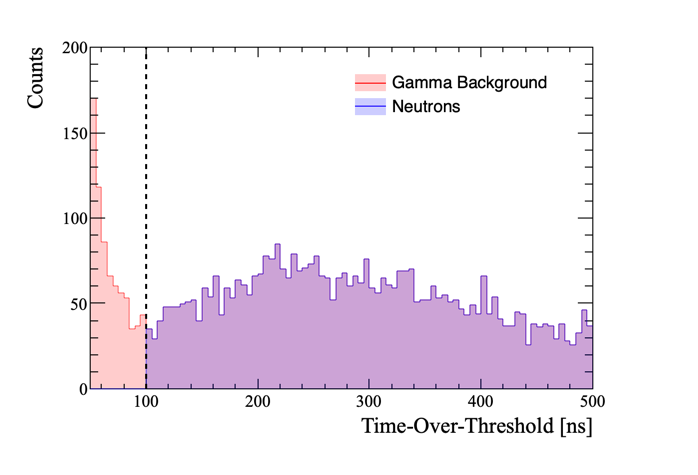}
    \caption{PSD distributions compared between a Cf-252 test source and a background testing case. (Top-left) Example pulse distributions for a gamma [blue] induced and neutron [red] induced event. (Top-right) PSD short/long distribution when the moderated detector was exposed to a Cf-252 source. (Bottom-right) PSD short/long distribution when the moderated detector was not exposed to a source. (Bottom-left) Time-over-threshold distribution obtained when the moderated detector was exposed to a Cf-252 source. }
    \label{fig:psdplots}
\end{figure}

Neutron pulses were discriminated from background by using software-based pulse shape discrimination. Events were first triggered in hardware on the DRS4 with a trigger threshold of -50~mV, before being transferred to the Raspberry Pi for software processing. A Pulse Shape Discrimination (PSD) metric was calculated from two gated integrals. The "short" integral was defined as the total pulse area below -5~mV and between -5~ns and +25~ns relative to the trigger time. The "long" integral was defined as the total pulse area below -5~mV and between +25~ns and +800~ns. The PSD metric was then defined as the ratio of short/long integrals. The distribution of PSD metric for different short pulse integrals is shown in Figure \ref{fig:psdplots} for a 24 hour exposure when operating the detector outside. A neutron PSD trigger was issued for any short/long ratio greater than 2. The timing windows and PSD trigger threshold were optimised based on initial tests with and without a high activity Cf-252 source. 

Example "neutron" and "background" pulses identified using this PSD metric are shown in Figure \ref{fig:psdplots}. Initial lab tests using Cf-252 were performed to assess the ability to detect neutrons using the PSD metric. Following an analysis of different pulse shape metrics it was observed that, since the dominant detector background was from much shorter pulses, it was also possible to select neutrons in the top region of the PSD plots shown in Figure \ref{fig:psdplots} using a simpler Time-Over-Threshold (TOT) cut applied to the digitized data in software. This alternative trigger condition is adopted in all of the software trigger analyses for the rest of this paper as it represents a trigger condition that can be realised with a simple single channel hardware discriminator. In future work optimisation of a triggering system to select high TOT \lifzns~ pulses entirely in hardware is planned, this will produce a low cost and low power consumption triggering system suitable for remote cosmic ray monitoring.

\section{Online Monitoring }
\label{sec:onlinemonitor}

\begin{figure}
    \centering
    \includegraphics[width=1.0\textwidth, clip, trim=0 100 0 0]{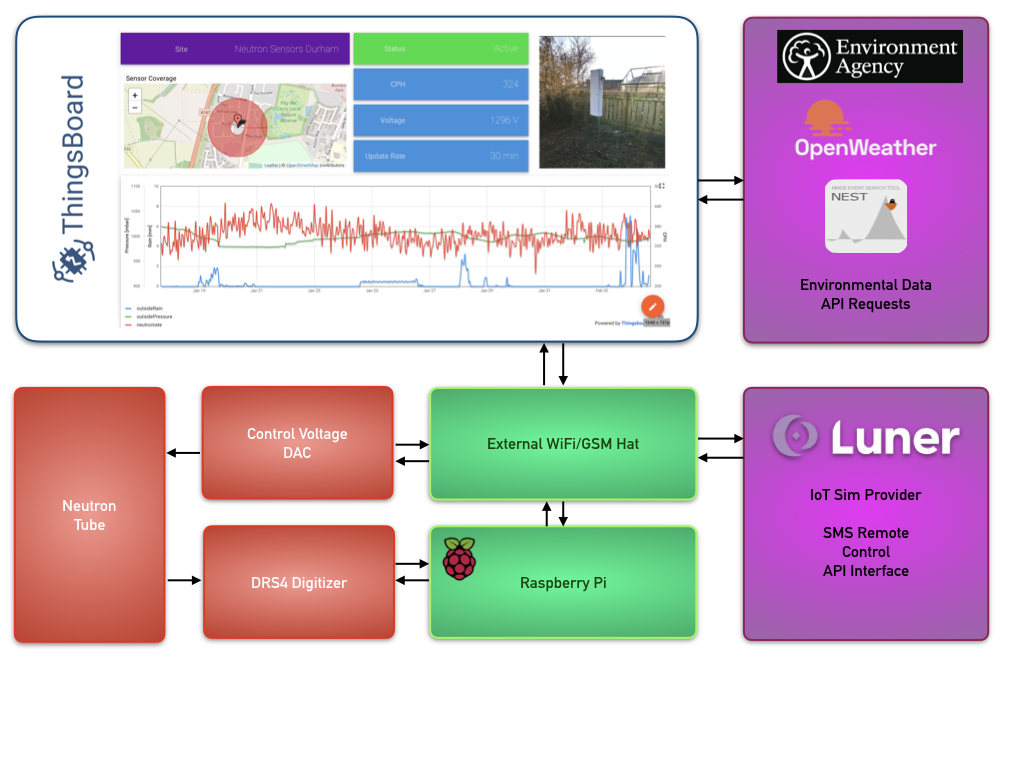}
    \caption{Remote neutron monitoring interface. Data processing is handled server side wherever possible using the open source Thingsboard IoT interface. This allows correlations to be extracted with respect to available weather information for the site, and reduces power consumption by offloading any required data corrections to the remote server.}
    \label{fig:thingsboard}
\end{figure}

To verify the assembled neutron detector was suitable for the monitoring of cosmic ray neutrons, long term monitoring data was taken over a period of two weeks with the detector installed outdoors in a residential environment in Durham, UK. The Raspberry Pi/DRS4 based data acquisition system described in the previous section was extended with a GSM interface hat to allow data to be transferred via 2G to a remote server for further processing. A Luner IoT SIM card was used to provide a network connection, allowing a simple control interface to be built upon Luner's existing  IoT communication API, enabling run settings such as high voltage or threshold settings to be configured remotely via SMS. Periodic data on local rainfall, pressure, and humidity was obtained automatically via the Thingsboard interface from both the Environment Agency and OpenWeather API. Similarly historic cosmic ray intensity data was obtained at midnight each day from the NEST neutron monitoring database \cite{mavromichalaki2011applications} to allow for corrections in the incoming cosmic ray intensity. A diagram of the deployment remote monitoring architecture is shown in Figure   \ref{fig:thingsboard}.

\begin{figure}
    \centering
    \includegraphics[width=0.49\textwidth, clip, trim=10 30 20 0]{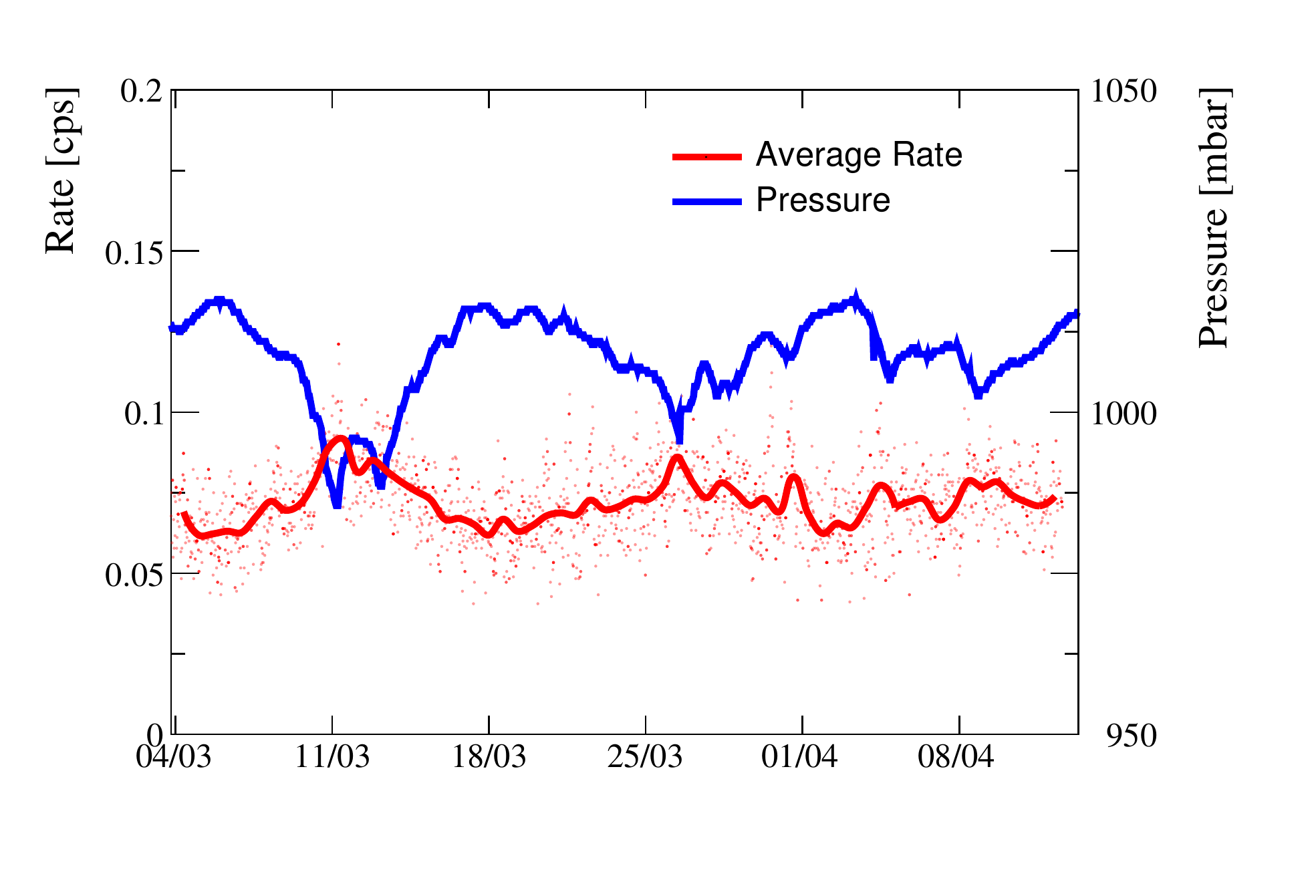}
    \includegraphics[width=0.49\textwidth, clip, trim=10 30 20 0]{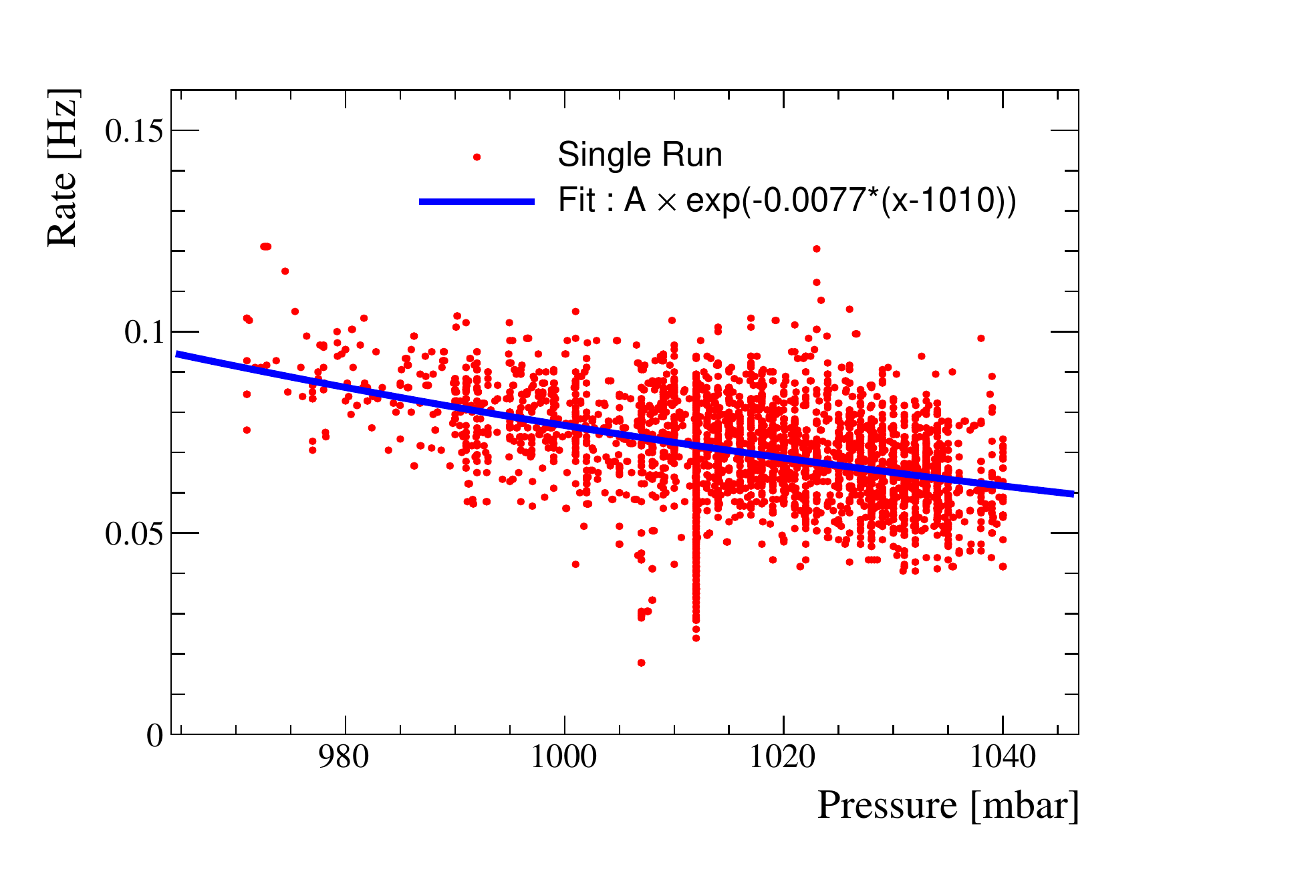}
    \caption{Measured system rate variation over a one month test deployment in a residential area. (left) Measured rate and pressure for each run as a function of date. (right) Correlation between pressure and rate.}
    \label{fig:finalrate}
\end{figure}

Over a period of approximately one month the detector system was found to achieve a good correlation with local pressure variations. As shown in Figure \ref{fig:finalrate} these variations were found to agree with the expected exponential drop in the neutron rate with increasing pressure. A small non-varying background rate, of approximately 0.03-0.04 Hz was observed during the deployment which was insensitive to the addition of HDPE or shielding materials. This is believed to be due to radon build-up within the scintillator assembly itself over the course of the test deployment. At present this background is quantified and subtracted using an exponential fit to the data as a function of pressure, assuming a fixed contribution that can be scaled up or down freely alongside the expected counting rate at 1010 mbar to achieve a good exponential fit to the data. The final trigger rate after this correction was found to be  approximately 3-4 times lower than expected for a CRS1000 detector, however the overall detector was assembled for approximately 10\% of the cost Helium-3 based system making it applicable for challenges where cost is a higher driving factor compared to temporal resolution. Work is ongoing to develop a low power trigger system that can effectively reduce this background contribution and improve the system's signal to noise ratio.

\section{Conclusions}
\label{sec:conclusion}
It has been demonstrated that scintillating thermal neutron detectors based upon \lifzns~or \bnzns~scintillator composites developed within the nuclear sector are viable solutions for neutron detection in the region of interest for soil moisture monitoring. Simulations of capture efficiency using GEANT4 found no difference between planar and cylindrical detector geometries when using a fixed volume of scintillator.

Simulations of light propagation inside internal light guides showed that wavelength shifting light guides are necessary when assembling cylindrical detectors longer than 5-10~cm, and that thinner rods were preferred when using wavelength shifting light guides. Low cost fluorescent light guides sourced from commercial suppliers were found to have adequate performance when producing smaller detectors, however extension of the proposed detector geometry beyond 50~cm would likely require custom manufactured rods with an increased absorption length in the 450-500~nm region.

Finally, the successful construction and deployment of a low-cost cylindrical neutron detector system was demonstrated. The system successfully operated without the requirement for additional fast neutron detectors or high sample digitiser readout electronics. A simple TOT cut was seen to deliver a similar performance for neutron discrimination compared to a full PSD integral analysis. Based on neutron triggers using a TOT > 100ns an approximate correlation between neutron counting rate and local pressure data has been demonstrated, this represents a promising first step in the development of ultra-low cost thermal neutron scintillator detectors for agricultural applications. In the future it is planned to co-deploy these systems alongside an array of commercially available in-ground based soil moisture probes to quantify, with varying choices of thermal neutron shielding materials, both the system performance and the effective detector footprint. 

\section*{Acknowledgements}
P.~Stowell would like to thank the Royal Commission for the Exhibition of 1851 for supporting this work. L.F.~Thompson and C. Steer would also like to thank the STFC for an earlier Opportunities call which laid the groundwork for this new detector design. P. Stowell would also like to thank P. Chadwick, C. Graham and C. Bourgenot for their guidance and support when measuring emission spectra of light guide geometries.

\bibliographystyle{unsrt} 
\bibliography{main}

\end{document}